\documentstyle[prl,aps,epsf,12pt]{revtex}
\begin{document}

\title{A Progress Report on the SO(5) Theory of High $T_c$ 
Superconductivity} 

\author{Shou-Cheng Zhang}
\address{\it Department of Physics, Stanford University, Stanford
  CA~~94305} 

\maketitle

\vspace{15pt}

\begin{abstract}
In this talk I give a brief update on the recent progress in the
$SO(5)$ theory of high $T_c$ superconductivity\cite{so5}. 
Reviewed topics include
$SO(5)$ ladders, the unification of BCS and SDW quasi-particles 
in the $SO(5)$ theory and the microscopic origin of the condensation 
energy.
\end{abstract}

\vspace{15pt}

First of all I would like to thank the 
Taniguchi foundation and the organizers of the Grand Finale
Taniguchi Symposium for inviting me to this wonderful conference.
This Symposium is appropriately entitled 
``The Physics and Chemistry of Transition Metal Oxides",
and it covers a vast and
broad range of topics whose intimate relations remain to be discovered.
In order to achieve a coherent grand synthesis in this subject, we 
must first
overcome the language barrier which separates
workers in different sub-fields.
At the welcome party of the Symposium, Professor George Swatzsky and I
were casually chatting about the $SO(5)$ theory of high $T_c$ 
superconductivity. A distinguished chemistry professor looked more
and more puzzled as he overheard our conversation. Finally, he couldn't
hold his curiosity and asked 
``$SO_5$? I didn't know that sulfer-pentoxide is a 
superconductor!"

In a series of conference proceedings, I tried to give a on-going
update about the status of the $SO(5)$ theory of HTSC\cite{m2s,sns}. 
This is the third one in this series. In the mean time, Auerbach
wrote a pedagogical review\cite{auerbach} explaining the $SO(5)$ theory in 
terms of more familiar concepts in $SO(3)$ quantum magnetism, and 
Hanke {\it et al} wrote a extensive review\cite{hanke-review} on the
numerical calculations within the $SO(5)$ theory.
Currently, the $SO(5)$ 
approach to HTSC is actively being investigated by many groups, 
focusing both on the microscopic origins and phenomenological
consequences.  Much progress has been made in understanding the
logical structure and examining the internal consistency of the theory.
In this report, I would first like
to summarize recent developments in understanding the microscopic
realization of the $SO(5)$ symmetry by using ladder systems as a theoretical 
laboratory\cite{szh,shelton,arrigoni,so8,duffy,eder2,schoutens,salk,furusaki,schulz}
and address the unification of BCS and SDW 
quasi-particles in the $SO(5)$ theory\cite{holonomy}. 
These developments reveal a
fascinatingly rich internal structure of the $SO(5)$ theory and could
ultimately lead to a microscopic foundation of the theory.
Although the $SO(5)$ theory appears to be a natural
framework to understand many experiments of HTSC in a unified 
fashion, no experiment has directly tested the fundamental validity
of the theory. Some more tests have been recently proposed in
\cite{bruus,burgess3}. However, I would like to focus on some recent 
works\cite{scalapino3,condensation} concerning
the microscopic origin of the condensation energy in HTSC, which could
lead to a direct and quantitative understanding of the microscopic
mechanism of HTSC. 

Shortly after the $SO(5)$ proposal, various groups have constructed
microscopic models with exact $SO(5)$ symmtry\cite{rabello,henley,burgess2}.
However, all this models
involve long range interactions which are not familiar and natural. 
It appears that as long as there is only one orbital per unit cell, this 
problem is unavoidable. For this reason the $SO(5)$ symmetry
was investigated for the two-legged ladder system, which has two
orbitals per unit cell, so that long ranged interaction can be 
avoided\cite{szh}.
Another reason for investigating $SO(5)$ symmetry in 
the ladder system is because it is a example of a Mott insulator without
any long range order
at half-filling. $SO(5)$ symmetry was original proposed as a theory
to unify antiferromagnetism (AF) with superconductivity (SC). It would be a
interesting question to see how it applies to Mott insulators without
any (quasi-) AF long range order. The third reason for investigating the
$SO(5)$ symmetry in the ladder system is to address the question of how 
this symmetry could emerge at long wave length without being present at
the microscopic level\cite{shelton,arrigoni,so8,schulz}. 
Because of quasi-one-dimensionality, well
controlled weak coupling RG calculations can be performed to address this
issue.

The fundamental quantity in the microscopic $SO(5)$ models is the concept
of a $SO(5)$ spinor\cite{rabello}, which has four components. 
On a two-legged ladder,
one could naturally combine the two sites on a rung to form such a 
spinor, and $SO(5)$ invariant models can be easily constructed by 
writing down the most general invariant interactions\cite{szh}. 
The parameter space for $SO(5)$ models is surprisingly large. Among the usual 
five local parameters $t$, $t_\perp$, $U$, $V$ and $J$, only one condition
is required to satisfy the $SO(5)$ symmetry ($U+V=J/4$). The Mott insulating
state at half-filling is not only a total spin singlet, but also a 
$SO(5)$ singlet. The lowest energy excitation on top of this singlet 
ground state is a five fold degenerate manifold of triplet magnons and
Cooper pairs. A uniform magnetic field or chemical potential can lower
the energy for one of these bosons, leading to a condensate with 
AF or SC quasi-long-range order. Within this framework, Mott insulator,
AF and SC states are intimately related and can be understood in a 
unified way. Many theoretical ideas about $SO(5)$ symmetry
can be tested in the ladders system. The eigenstates of the $SO(5)$
ladder models can all be classified into general irreducible representations
of the $SO(5)$ group. These states form a beautiful and revealing pattern
and have been identified in numerical calculations by
Eder, Dorneich, Zacher, Hanke and the author\cite{eder2}.
The photoemission spectrum of the $SO(5)$ ladder has been studied in 
details\cite{eder2,salk}. The single electron Green's functions can be 
related by exact Ward identities and are shown by 
direct diagonalization to evolve continuously from
the Mott insulator to the superconducting state\cite{eder2}. 
The structure of the exact $\pi$ resonance can also be understood 
in detail both analytically
and numerically\cite{furusaki,eder2}. Due to the quasi-long-range 
order of the superconducting
state, the $\pi$ resonance is not a delta function peak, but a 
threshold singularity at energy $-2\mu$\cite{szh,furusaki}.
Other physical quantities can also be calculated\cite{pivovarov}.

While the $SO(5)$ models offer a nice testing ground for many interesting
theoretical ideas, the parameters of these models are not
``realistic". It is therefore desirable to understand whether they share
some common features with more realistic ladder models. For example,
in the strong coupling limit, the ground state of a generic ladder model 
is a product state of singlet rungs. 
Rather surprisingly, this state is not only a total spin singlet,
but also a $SO(5)$ singlet, since this product state is annihilated
by the $\pi$ operators. In any $SO(5)$ singlet state, the behavior
of the static AF and SC correlation functions are identical. Therefore, one
would expect these correlations to scale towards a common behavior in the
strong coupling limit of a generic ladder model. Unfortunately, this
argument does not apply to the dynamic correlation functions. 
One can use the exact $SO(5)$ models as a point of departure to
systematically vary the symmetry violating perturbations and compare
the results with the partial multiplet structures found in the
$t-J$ model\cite{eder}. Investigation towards this direction has been
taken in ref.\cite{duffy,eder2}.

Fortunately, in the weak coupling regime, one can perform controlled 
RG calculations to see how exact symmetries emerge from 
nonsymmetric interactions\cite{shelton,arrigoni,so8,schulz}. 
Lin, Balents and Fisher\cite{so8} showed that there are in 
general 9 marginal
operators for a two-legged ladder at half-filling, 5 of them preserve
$SO(5)$ symmetry, while 4 of them violate it. Surprisingly, the 
symmetry violating interactions scale to zero. Within the remaining
$SO(5)$ manifold, there are four stable fix points, with even higher
symmetry, namely $SO(8)$. Arrigoni and Hanke\cite{arrigoni} also studied the
symmetry violating band structure effects, for example the 
next-nearst-neighbor hopping $t'$, and demonstrated that after a
suitable redefinition, the symmetry violating effects can be completely
absorbed. More recently, Schulz\cite{schulz} extended the RG calculations
from half-filling to general filling, and showed that $SO(6)$ and $SO(5)$
symmetries can emerge dynamically. These developments are very exciting,
and offer hope that similar symmetry restoration effects can occur
in higher dimensional systems near a quantum critical 
point\cite{so5,laughlin}.

The original $SO(5)$ theory was formulated purely in terms of bosonic
collective degrees of freedom\cite{so5}. 
It is clear that a complete theory has
to include the fermionic sector. The bosonic theory explains how various
collective modes connect to each other at the transition between
the AF and SC states, it is natural to ask how the fermionic BCS and
SDW quasi-particles connect to each other. The answer to this question
turns out to be surprisingly rich and beautiful. In elementary
quantum mechanics, we learned about a remarkable demonstration of the
spinor nature of the electron. If two strong magnetic fields
polarize the electron spins in two orthogonal directions, 
then the electron
wave functions corresponding to these two orthogonal fields are not
orthogonal to each other. This non-orthogonality of the electron wave
functions allows transmission of a electron beam through the
orthogonal field regions. It is also the origin of the Berry's phase. 
Quite similarly, even though
a antiferromagnet and a superconductor have ``orthogonal" order parameters,
the quasi-particles associated with these two states are not orthogonal
to each other. This non-orthogonality leads to a novel generalization of
Berry's phase to a $SU(2)$ holonomy\cite{holonomy}. The generalization
from the $U(1)$ Berry's phase of a $SO(3)$ spinor to the $SU(2)$ holonomy
of a $SO(5)$ spinor is a unique generalization in a precise mathematical
sense, and involves some of the most beautiful, and yet seeming disconnected
mathematical concepts such as Hopf maps, quaternions and the Yang monopole.
So far this work is still at a mathematical stage, and the precise
physical implications can only be anticipated at this moment. Potential
applications include a generalized Bogoliubov-deGennes type of formalism
to discuss novel fermionic excitations near topological defects involving 
twists of the $SO(5)$ superspin 
vector\cite{vortex,junction1,junction2,goldbart1},
a novel type of Andreev reflection
at the AF/SC boundary, a non-abelian Bohm Aharonov effect associated with
regions with non-trivial superspin twists and novel understanding of the
single particle properties in the pseudogap regime. The remarkablly rich
fermionic structure in the $SO(5)$ model shows that the $SO(5)$ theory is
much more than a expanded version of the Landau-Ginzburg theory, it can 
fully address single particle excitations and their coupling to the 
collective modes.

The ideas on the $SU(2)$ holonomy can possibly be extended to other physical
systems as well, especially transition metal oxide systems with
orbital degeneracy. Soon after the discovery of quantum mechanics,  
Wigner and von Neumann studied the problem of generic level crossings
in quantum mechanics. They and later Dyson classified generic level
crossings into three categories, now called the orthogonal, unitary
and symplectic ensembles. The familiar $U(1)$ Berry's phase occur in the
unitary ensemble. In the sympletic ensemble, one deals with time reversal
invariant systems with Kramers degeneracy. The level crossing between
two Kramers doublets can be described by the four dimensional (since
two doublets give four states in total) symplectic group $Sp(4)$, which
happens to be isomorphic to $SO(5)$. This type of level crossing phenomenon
can not only occur at AF/SC transition, but also in generic problems
involving spin-orbit couplings. The ideas on $SU(2)$ holonomy can not only
lead to deeper understandings of these systems, but may also help to 
understand their (formal) relationship to the high $T_c$ problem.

Although the $SO(5)$ theory was originally proposed as a effective theory to
understand the interplay between AF and SC, it implicitly points to a
microscopic mechanism for HTSC. Within this theory, the microscopic
mechanism for SC is basically the ``same" as the microscopic mechanism
for AF, namely the lowering of the exchange energy 
$J\sum_{i,j} \vec S_i \vec S_j$. Recently, Scalapino and 
White\cite{scalapino3} argued that
the lowering of the exchange energy can be quantitatively correlated with
the SC condensation energy. This insightful observation allows for
quantitative test of various mechanisms of HTSC which should make detailed
prediction on how the exchange energy is saved in the SC state. The $SO(5)$
theory predicts a $\pi$ resonance mode\cite{so5,resonance1,resonance2}
which is identified with the 
neutron resonance mode observed in the SC state. The theory
of the neutron resonance mode is based on a particle particle collective
mode near momentum $(\pi,\pi)$, which exists both in the normal and
SC state. Since a particle particle mode can only make a contribution
to the spin correlation function in the SC state, the neutron resonance
mode is observed only below $T_c$. Recently, Demler and I 
noticed\cite{condensation} that
this argument also provides a concrete microscopic mechanism for HTSC. 
It is straightforward to see that the coupling to a particle particle 
collective mode around $(\pi,\pi)$ gives a {\it negative} difference
between the exchange energy $J\sum_{i,j} \vec S_i \vec S_j$ in the
SC and the normal state. Therefore, the SC saves more exchange energy
compared to the normal state, and the amount of saving is precisely
given by $J$ times the (dimensionless) integrated spectral weight of the
$\pi$ resonance. From the neutron scattering experiments by 
Fong et al\cite{keimer2},
one can see that the change in the dimensionless quantity
$\vec S_i \vec S_j$ due to the $\pi$
resonance is on the order of few per cent, which gives a saving of
exchange energy of $35K$ per unit cell. On the other hand, the condensation
energy of optimally doped $YBCO$ superconductor is about $5K$\cite{loram}. 
Therefore,
we see that the emergence of the $\pi$ resonance could be the dominant
mechanism responsible for the superconducting condensation energy.
Upon going to the SC state, the kinetic energy usually increases, so that
the saving in exchange energy could be balanced by the cost in kinetic
energy to give the right condensation energy. It is therefore highly
desirable to find direct ways to measure the change in kinetic energy
when the system enters the SC state.

This line of reasoning leading to a microscopic mechanism of SC is 
somewhat unfamiliar, since most people equate the SC mechanism with
a form of attractive interaction between electrons. However, I would
like to argue that in a strongly correlated system,
this new line of thinking is much more fruitful and experimentally 
accessible. {\it The central idea here is to identify a energy saving
process which is forbidden in the normal state but possible in the
SC state}. In our mechanism, the particle particle resonance is just 
such a process. The only other example I can think of is the interlayer
tunneling mechanism\cite{ILT}. 
In this case, the energy saving process is the
c axis tunneling, which is forbidden in the normal state if the normal
state is not a fermi liquid, but is allowed in the SC state. In both
examples we see that once such a process is identified, experimentally
falsifiable prediction about the condensation energy follow immediately.
Following this line of thinking, we can hopefully move the debate about
the microscopic mechanism of HTSC to a new level, where direct
comparison with experiments becomes possible.

Since this argument seem to strongly rely on the onset of the $\pi$
resonance at $T_c$ for the optimally doped superconductors, a alert
reader may wonder how this argument applies to the underdoped 
superconductors, where a broadened resonance peak is observed above
$T_c$ but below the pseudogap temperature $T_{MF}$\cite{mook1,keimer3}. 
Let us first see
how the $SO(5)$ theory could explain the broadened peak in the
pseudogap regime. The basic process in the SC state is given by the
following Feymann diagram.

\begin{figure*}[h]
\centerline{\epsfysize=3.8cm 
\epsfbox{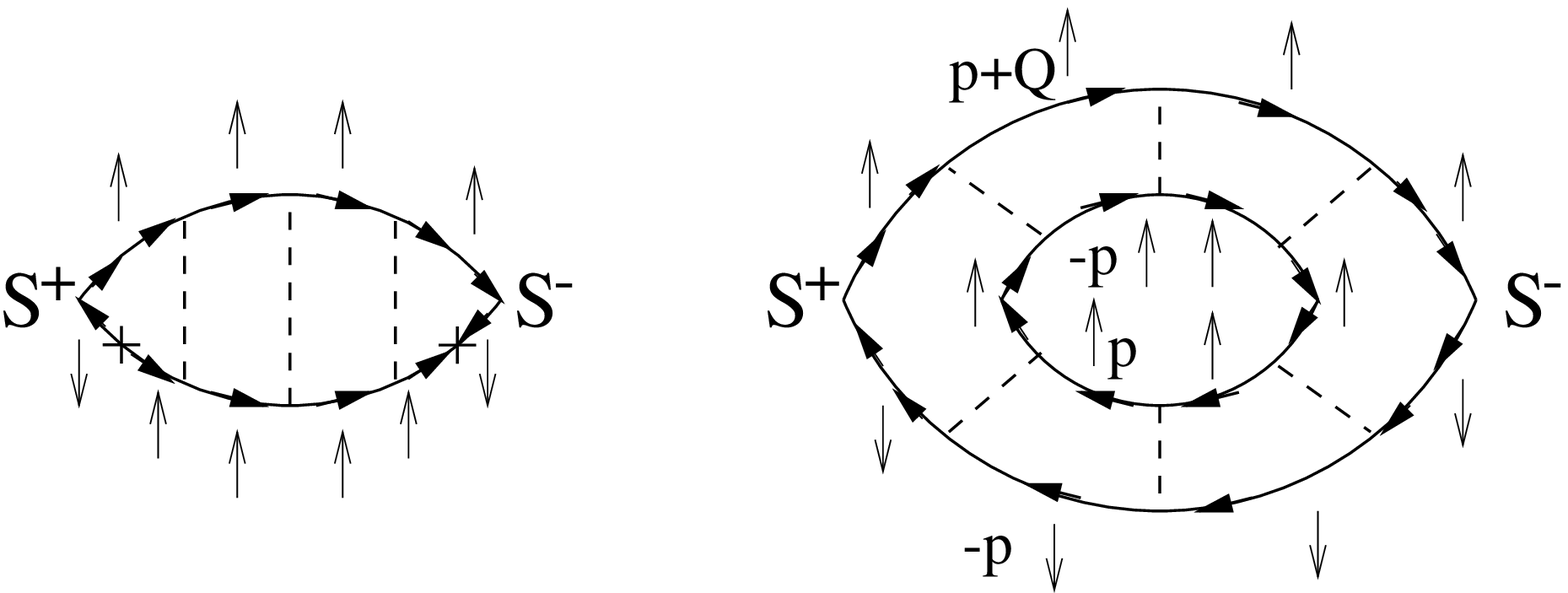}
}
\caption{Feymann diagram for the $\pi$ resonance in the SC (left panel)
and normal (right panel) state. In the SC state, a particle hole pair
created by the neutron is converted into a particle particle pair
by the SC condensate (marked by two crosses). In the normal state,
such a anomalous process is forbidden, but one can open up the crosses
and reconnect them to obtain a Cooper pair propagator. In the pseudogap
regime, there is no sharp pole due to a single $\pi$ pair, but 
a broadened convolution of a $\pi$ and a Cooper pair.}
\label{diagram}
\end{figure*}

We see that the spin vertex creates a 
particle hole pair, but the hole can be converted into a particle
by the Gorkov $F$ function, and the multiple scattering in the
particle particle channel gives rise to a sharp collective mode in
the dynamic spin correlation function. In the normal state, the
Gorkov $F$ function vanishes identically so that such a process is
not possible. However, one could cut two Gorkov $F$ functions and
clue them together in the normal state as shown in Fig 1. Such a
process does not vanish in the normal state, and
represents the preformed Cooper pair fluctuations in the pseudogap
regime. In this case, rather than a single resonance peak, we obtain
a convolution between the triplet momentum $(\pi,\pi)$
particle particle resonance and the singlet zero momentum Cooper
pair resonance. If these resonances are weakly dispersive, the convolution
spectrum will be reasonably sharp. Since the basic processes below
and above $T_c$ are different, we would generally expect some
discontinuous behavior at $T_c$. The experimental plots of the intensity
as a function of temperature does appear to consist of two separate
curves joining with a discontinuous derivative at $T_c$.
Having understood the SC fluctuation in the normal state as the origin
of the neutron intensity above $T_c$, we would use the general arguments
outlined above to correlate the neutron scattering intensity with the
condensation energy in the pseudogap regime. Remarkably, the 
condensation energy obtained from Loram's specific data (Fig. 8 of
reference \cite{loram}) and $\pi$ resonance intensity measured in
neutron scattering\cite{mook2} in the 
underdoped regime have the same qualitative behavior, namely consisting
of two separate curves joining with a discontinuous derivative at $T_c$.
The remarkable similarity between these two seemingly different experiments
lend strong support to our interpretation, and could not only lead to
a quantitative understanding of the microscopic mechanism of HTSC, but also
the origin of the pseudogap physics. 

I would to thank E. Demler, R. Eder, A. Furusaki, W. Hanke,    
S. Rabello and D. Scalapino for close collaborations on projects reported 
above. 
This work is supported by the NSF under grant numbers DMR-9400372 
and DMR-9522915.


\end{document}